\documentclass[12pt]{article}
\usepackage{amssymb}
\newcommand{\be}{\begin{eqnarray}}
\newcommand{\ee}{\end{eqnarray}}
\newcommand{\non}{\nonumber}
\newcommand{\SSS}{\ensuremath{\mathsf{S}}}

\begin{document}

\begin{titlepage}
\strut\hfill UMTG--227
\vspace{.5in}
\begin{center}

\LARGE The boundary supersymmetric sine-Gordon model revisited\\[1.0in]
\large Rafael I. Nepomechie\\[0.8in]
\large Physics Department, P.O. Box 248046, University of Miami\\[0.2in]  
\large Coral Gables, FL 33124 USA\\

\end{center}

\vspace{.5in}

\begin{abstract}
We argue that, contrary to previous claims, the supersymmetric 
sine-Gordon model with boundary has a two-parameter family of boundary 
interactions which preserves both integrability and supersymmetry.  We 
also propose the corresponding boundary $S$ matrix for the first 
supermultiplet of breathers.
\end{abstract}

\end{titlepage}

\setcounter{footnote}{0}

\section{Introduction}\label{sec:intro}

Ghoshal and Zamolodchikov \cite{GZ} formulated the general framework
of integrable quantum field theory in the presence of a boundary. 
\footnote{Earlier related work includes \cite{Ch}-\cite{FK}.}
Among the examples which they investigated is the boundary version of
the sine-Gordon (SG) model \cite{ArKo, ZZ}.  They argued that this
model has a two-parameter family of boundary interactions which
preserves integrability, and they proposed the corresponding soliton
boundary $S$ matrix.

Inami {\it et al.} \cite{IOZ} subsequently investigated a boundary
version of the supersymmetric sine-Gordon (SSG) model
\cite{ssg1}-\cite{ssg3}.  They argued that the combined constraints of
integrability and supersymmetry do not allow {\it any} free parameters
in the boundary interaction.  The boundary SSG and SShG
(supersymmetric sinh-Gordon) models have been studied further in
\cite{AK}-\cite{AC}.

We argue here that, contrary to the claim in \cite{IOZ}, the boundary
SSG model has a two-parameter family of boundary interactions which
preserves both integrability and supersymmetry.  The key point is that
the class of boundary interactions considered in \cite{IOZ} is too
restrictive: one must instead introduce a Fermionic boundary degree of
freedom, as was done by Ghoshal and Zamolodchikov \cite{GZ} in order
to describe the Ising model in a boundary magnetic field.  The
boundary SG model at the free Fermion point \cite{AKL, MN} is also
formulated with such Fermionic boundary degrees of freedom.

Our result resolves an interesting paradox which arose from our
recent work \cite{AN2} on the boundary supersymmetric Yang-Lee model,
which is defined using the notion of perturbed conformal field theory
\cite{Za}.  Indeed, let us first recall that, just as the bulk
Yang-Lee (YL) model \cite{CM} is a restriction \cite{Sm} of the bulk
SG model, the bulk supersymmetric Yang-Lee (SYL) model \cite{Sc} is a
restriction \cite{Ah1} of the bulk SSG model.  The same is true for
the corresponding boundary versions: the boundary YL \cite{GZ, DPTW}
and boundary SYL \cite{AN2} models are restrictions of the boundary SG
and boundary SSG models, respectively.  In \cite{AN2} we found strong
evidence (both from perturbed conformal field theory and the
conjectured boundary $S$ matrix) that the boundary SYL model with one
free boundary parameter has supersymmetry.  This strongly suggests
that its ``parent'', the boundary SSG model, should also have at least
a one-parameter family of boundary interactions which maintains both
integrability and supersymmetry.

The outline of this Letter is as follows.  In Section 2, we propose 
the boundary interaction in terms of two unknown functions $f(\phi)$ 
and ${\cal B}(\phi)$.  We determine these functions in Section 3 from 
the requirements of supersymmetry and integrability.  In Section 4 we 
briefly discuss our results.  In particular, we propose the boundary 
$S$ matrix for the first supermultiplet of breathers.

\section{The boundary SSG model}\label{sec:SSG}

In order to facilitate comparison with \cite{IOZ}, which hereafter we
refer to as I, we adopt the same notations.  The Euclidean-space
action of the boundary SSG model is given by
\be
S &=& \int_{-\infty}^{\infty} dy \int_{-\infty}^{0} dx\ {\cal L}_{0}
+ \int_{-\infty}^{\infty} dy\  {\cal L}_{b} \,,
\ee
where the bulk Lagrangian density is given by\footnote{We have already
performed, following I, the field rescalings $\phi \rightarrow
\phi/\beta$, $\psi \rightarrow \psi/\beta$, $\bar \psi \rightarrow
\bar\psi/\beta$, and we consider the classical limit $\beta
\rightarrow 0$; moreover, we have set the mass $m=2$ (cf.  Eq. 
(I2.1)).}
\be
{\cal L}_{0} =
2 \partial_{z}\phi \partial_{\bar z} \phi  
- 2 \bar \psi \partial_{z} \bar \psi 
+ 2 \psi \partial_{\bar z} \psi 
- 4 \cos \phi - 4 \bar \psi \psi \cos {\phi\over 2} \,,
\label{bulkL}
\ee
where $\psi$ and $\bar \psi$ are the two components of a Majorana
Fermion field, and $z=x+iy$, $\bar z=x-iy$.  We propose the following
boundary Lagrangian at $x=0$
\be
{\cal L}_{b} = \bar \psi \psi + ia \partial_{y} a 
- 2 f(\phi) a (\psi - \bar \psi) + {\cal B}(\phi) \,,
\label{boundL}
\ee
where $a$ is a Hermitian Fermionic boundary degree of freedom which
anticommutes with both $\psi$ and $\bar \psi$.  As mentioned in the
Introduction, such an approach was first considered by Ghoshal and
Zamolodchikov \cite{GZ} to describe the Ising model in a boundary
magnetic field.  Moreover, $f(\phi)$ and ${\cal B}(\phi)$ are
potentials (functions of the scalar field $\phi$, but not of its
derivatives), which are still to be determined.

Variation of the action gives the classical equations of motion. For 
the bulk, the equations are (I2.2), (I3.3)
\be
\partial_{\bar z} \partial_{z} \phi &=& 
\sin \phi + {1\over 2} \bar \psi \psi \sin {\phi\over 2}\,, \non \\
\partial_{\bar z} \psi &=&
- \bar \psi \cos {\phi\over 2} \,, \qquad 
\partial_{ z}\bar \psi = - \psi \cos {\phi\over 2} \,,
\label{bulkeom}
\ee 
and the boundary conditions at $x=0$ are
\be
\psi + \bar \psi &=& 2 f a \,, \label{boundeom1} \\
i\partial_{y} a &=& f (\psi - \bar \psi) \,, \label{boundeom2} \\
\partial_{x} \phi &=& -{\partial {\cal B}\over \partial \phi} 
+ 2 {\partial f\over \partial \phi} a (\psi - \bar \psi) \,.
\label{boundeom3}
\ee
If we assume that $f$ is nonzero, then we can eliminate $a$ from Eqs.  
(\ref{boundeom1})-(\ref{boundeom3}), and thereby obtain the following 
boundary conditions at $x=0$:
\be
\partial_{y}(\psi + \bar \psi)  &=& 
{\partial \ln f\over \partial \phi} \partial_{y} \phi (\psi + \bar \psi)
- 2 i f^{2} (\psi - \bar \psi) \,, \label{bcF} \\
\partial_{x} \phi &=& -{\partial {\cal B}\over \partial \phi} 
+ 2 {\partial \ln f\over \partial \phi} \bar \psi \psi \,.
\label{bcB}
\ee

\section{Integrals of motion}

We now proceed to determine the potentials $f(\phi)$ and ${\cal 
B}(\phi)$ which appear in the boundary action by demanding both 
supersymmetry and integrability.  We recall \cite{IOZ, ssg2} that the 
bulk SSG model has an infinite number of integrals of motion 
constructed from the densities $T_{s+1}$, $\overline{T}_{s+1}$, 
$\Theta_{s-1}$, $\overline{\Theta}_{s-1}$, which obey
\be
\partial_{\bar z} T_{s+1} = \partial_{z} \Theta_{s-1} \,, 
\qquad 
\partial_{z} \overline{T}_{s+1} 
= \partial_{\bar z} \overline{\Theta}_{s-1} \,.
\label{continuity}
\ee 
The $s={1\over 2}$ and $s=1$ densities are given by the supercurrent
\be
T_{3\over 2} &=& \partial_{z}\phi \psi \,, \qquad
\overline{T}_{3\over 2} = \partial_{\bar z}\phi \bar \psi \,, \non  \\
\Theta_{-{1\over 2}} &=& -2 \bar \psi \sin {\phi\over 2} \,, \qquad
\overline{\Theta}_{-{1\over 2}} = -2 \psi \sin {\phi\over 2} \,,
\ee
and the energy-momentum tensor
\be
T_{2} &=& (\partial_{z}\phi)^{2} - \partial_{z}\psi \psi\,, \qquad
\overline{T}_{2} = (\partial_{\bar z}\phi)^{2} 
+ \partial_{\bar z}\bar \psi \bar \psi \,, \non  \\
\Theta_{0} &=& \overline{\Theta}_{0} = -2 \cos \phi 
- \bar \psi \psi \cos {\phi\over 2} \,,
\ee
respectively. The $s=3$ densities are given by \cite{IOZ, ssg2}
\be
T_{4} &=& (\partial_{z}^{2}\phi)^{2} - {1\over 4}(\partial_{z}\phi)^{4}
+ {3\over 4} (\partial_{z}\phi)^{2} \partial_{z}\psi \psi 
- \partial_{z}^{2}\psi \partial_{z}\psi  \,, \non  \\
\overline{T}_{4} &=& (\partial_{\bar z}^{2}\phi)^{2} 
- {1\over 4}(\partial_{\bar z}\phi)^{4}
- {3\over 4} (\partial_{\bar z}\phi)^{2} \partial_{\bar z}\bar \psi \bar \psi 
+ \partial_{\bar z}^{2}\bar \psi \partial_{\bar z}\bar \psi  \,, \non  \\
\Theta_{2} &=& (\partial_{z}\phi)^{2} \cos \phi 
- \partial_{z}\psi \psi \cos^{2} {\phi\over 2}
- \bar \psi \partial_{z}\psi \partial_{z} \cos {\phi\over 2}
+ {1\over 4} \bar \psi \psi (\partial_{z}\phi)^{2} \cos {\phi\over 2} 
\,, \non  \\
\overline{\Theta}_{2} &=& (\partial_{\bar z}\phi)^{2} \cos \phi 
+ \partial_{\bar z}\bar \psi \bar \psi \cos^{2} {\phi\over 2}
+ \psi \partial_{\bar z}\bar \psi \partial_{\bar z} \cos {\phi\over 2}
+ {1\over 4} \bar \psi \psi (\partial_{\bar z}\phi)^{2} \cos {\phi\over 2} 
\,.
\ee

As observed by Ghoshal and Zamolodchikov \cite{GZ}, it follows from
the continuity Eqs.  (\ref{continuity}) that the boundary model has
the integral of motion
\be
P_{s}= \int_{-\infty}^{0}dx\ (T_{s+1} + \overline{T}_{s+1} + 
\Theta_{s-1} + \overline{\Theta}_{s-1}) - i\Sigma_{s}(y) \,,
\label{iom}
\ee
provided that the following condition holds at $x=0$
\be
T_{s+1} - \overline{T}_{s+1} - \Theta_{s-1} + \overline{\Theta}_{s-1}
= \partial_{y} \Sigma_{s}(y) \,,
\label{constraint}
\ee 
where $\Sigma_{s}(y)$ is a (local) boundary term.

As we shall show below, the requirement that $P_{3}$ be an integral of 
motion implies that the potential ${\cal B}(\phi)$ is the same as for 
the (Bosonic) boundary SG model \cite{GZ}
\be
{\cal B}(\phi) = 2 \alpha \cos {1\over 2}(\phi - \phi_{0}) \,,
\label{Bpotential}
\ee
where $\alpha$ and $\phi_{0}$ are arbitrary real parameters.

The requirement of on-shell supersymmetry is that $P_{1\over 2}$ also 
be an integral of motion.  The constraint (\ref{constraint}) for 
$s={1\over 2}$ reads
\be
% T_{3\over 2} - \overline{T}_{3\over 2} - \Theta_{-{1\over 2} +
% \overline{\Theta}_{-{1\over 2} =
({1\over 2}\partial_{x}\phi - 2 \sin {\phi\over 2})(\psi-\bar \psi)
- {i\over 2} \partial_{y}\phi (\psi+\bar \psi) =  \partial_{y} 
\Sigma_{1\over 2} \,.
\label{susyconstraint}
\ee
We now substitute for $\partial_{x}\phi$ the boundary condition 
(\ref{bcB}) with the potential ${\cal B}(\phi)$ given by 
(\ref{Bpotential}); and then we use (\ref{boundeom1}) and 
(\ref{boundeom2}) to substitute for $\psi+\bar \psi$ and $\psi-\bar 
\psi$, respectively.  Moreover, we make the following Ansatz for the 
boundary term
\be
\Sigma_{1\over 2} = {i\over 2} g(\phi) a \,,
\label{boundterm}
\ee 
where $a$ is the Fermionic boundary degree of freedom appearing in the
boundary Lagrangian (\ref{boundL}), and $g(\phi)$ is a function of
$\phi$ which is yet to be determined.  Eq.  (\ref{susyconstraint})
implies that $g(\phi)$ must satisfy the pair of equations
\be
g = {\alpha \sin{1\over 2}(\phi - \phi_{0}) 
- 4 \sin {\phi\over 2}\over f} \,, \qquad 
{\partial g\over \partial \phi} = -2 f \,.
\ee
It is not difficult to solve these equations for $f(\phi)$ (and hence,
also $g(\phi)$),
\be
f(\phi) = \left[ \alpha \sin{1\over 2}(\phi - \phi_{0}) 
- 4 \sin {\phi\over 2}\right]\left[  {1\over 8}{1\over
\alpha \cos{1\over 2}(\phi - \phi_{0}) 
- 4 \cos {\phi\over 2} + C}\right]^{1\over 2} \,,
\label{fresult}
\ee
where $C$ is an integration constant.

We now turn to the requirement that $P_{3}$ be an integral of motion. 
The LHS of the constraint (\ref{constraint}) for $s=3$ is given by
(I3.10)
\be
\lefteqn{T_{4} - \overline{T}_{4} - \Theta_{2} + \overline{\Theta}_{2}
= } \non  \\ 
& & -{i\over 2}( 4 \sin \phi - 2 \phi_{yy} + 2 \bar \psi \psi 
\sin{\phi\over 2}) \phi_{xy} 
+ {i\over 8}(\phi_{x}^{2}-\phi_{y}^{2})\phi_{x} \phi_{y} 
+ i \phi_{x} \phi_{y} \cos \phi \non  \\ 
& &+ {3i\over 16} (\phi_{x}^{2}-\phi_{y}^{2})
(\bar \psi_{y} \bar \psi - \psi_{y} \psi) 
+ 3i \cos^{2}{\phi\over 2} (\bar \psi_{y} \bar \psi - \psi_{y} \psi) 
+{i\over 2}\sin {\phi\over 2} \phi_{x}
(\bar \psi_{y} \psi - \psi_{y} \bar \psi) \non  \\ 
& &+ i(\bar \psi_{yy} \bar \psi_{y} - \psi_{yy} \psi_{y})
+i \bar \psi \psi \phi_{x} \phi_{y}  \cos {\phi\over 2} 
-{3\over 8}\phi_{x} \phi_{y} 
(\bar \psi_{y} \bar \psi + \psi_{y} \psi)\non  \\ 
& &+ 2 \partial_{y} \cos {\phi\over 2}
(\bar \psi_{y} \psi - \bar\psi \psi_{y})
+ \cos {\phi\over 2}(\bar \psi \psi_{yy} - \bar\psi_{yy} \psi) \,.
\label{integconstraint}
\ee
Our task is to try to write the above expression as $\partial_{y} 
\Sigma_{3}$.

We begin by replacing $\phi_{x}$ by the boundary condition 
(\ref{bcB}), and $\phi_{xy}$ by
\be
\phi_{xy} =  -{\partial^{2} {\cal B}\over \partial \phi^{2}} \phi_{y}
+  2 {\partial^{2} \ln f\over \partial \phi^{2}} \phi_{y}\bar \psi \psi 
+  2 {\partial \ln f\over \partial \phi} 
(\bar \psi_{y} \psi + \bar \psi \psi_{y}) \,.
\ee
Since the pure Bosonic terms in (\ref{integconstraint}) (which
evidently do not depend on the potential $f$) must by themselves form
a total $y$-derivative, it is clear that ${\cal B}(\phi)$ must be the
same as for the pure Bosonic boundary SG model, as anticipated in
Eq.  (\ref{Bpotential}) above.  (See also I.)

The last two terms in (\ref{integconstraint}) can be re-expressed as
\be
3 \partial_{y} \cos {\phi\over 2} (\bar \psi_{y} \psi - \bar \psi \psi_{y})
+ \partial_{y} \left[\cos {\phi\over 2}
(\bar \psi \psi_{y} - \bar\psi_{y} \psi) \right] \,.
\ee

Let us now consider the Fermionic term in (\ref{integconstraint}) with 
three derivatives, $i(\bar \psi_{yy} \bar \psi_{y} - \psi_{yy} 
\psi_{y})$.  We can extract a total $y$-derivative from this term as 
follows: first, observe the identity
\be
\partial_{y} (\psi_{y} \bar \psi_{y})= \psi_{yy} \bar\psi_{y} + 
\psi_{y}\bar \psi_{yy} \,.
\label{step1}
\ee 
Next, differentiate the boundary condition (\ref{bcF}),
\be
\lefteqn{\psi_{yy} + \bar \psi_{yy} = 
\left({\partial^{2}\ln f\over \partial \phi^{2}}  \phi_{y}^{2} 
+ {\partial \ln f \over \partial \phi} \phi_{yy} \right)(\psi + \bar 
\psi)}\non \\
& &+ {\partial \ln f\over \partial \phi} \phi_{y}(\psi_{y} + \bar \psi_{y})
-4 i f^{2} {\partial \ln f \over \partial \phi}\phi_{y}(\psi - \bar \psi) 
-2 i f^{2} (\psi_{y} - \bar \psi_{y}) \,.
\label{step2}
\ee
Solving (\ref{step2}) for $\psi_{yy}$ and substituting into 
(\ref{step1}), and repeating for $\bar\psi_{yy}$, we obtain the 
desired result
\be
\lefteqn{i(\bar \psi_{yy} \bar \psi_{y} - \psi_{yy} \psi_{y}) =
-i\partial_{y} (\psi_{y} \bar \psi_{y})} \non  \\
& &+i \left({\partial^{2}\ln f\over \partial \phi^{2}}  \phi_{y}^{2} 
+ {\partial \ln f \over \partial \phi} \phi_{yy} \right)
\Big[(\psi + \bar \psi)\bar \psi_{y} + \psi_{y}(\psi + \bar 
\psi)\Big] \non \\
& &+4 f^{2} {\partial \ln f \over \partial \phi}\phi_{y}
\Big[(\psi - \bar \psi)\bar \psi_{y} + \psi_{y}(\psi - \bar \psi)\Big]
+2i {\partial \ln f \over \partial \phi}\phi_{y} \psi_{y}\bar \psi_{y}
\,.
\ee 

Our general strategy now is to use the boundary condition (\ref{bcF}) 
to express $\psi_{y} \bar \psi$ in terms of $\bar \psi_{y} \bar \psi$ 
(plus terms with $\bar \psi \psi$); and similarly to express $\bar 
\psi_{y} \psi$ in terms of $\psi_{y} \psi$, and $\psi_{y}\bar 
\psi_{y}$ in terms of $\bar \psi_{y} \bar \psi$ and $\psi_{y} \psi$.  
In this way, all the remaining Fermion bilinears have at most one 
$y$-derivative; and those bilinears with one derivative appear in only 
two combinations: $(\bar \psi_{y} \bar \psi - \psi_{y} \psi)$ and 
$(\bar \psi_{y} \bar \psi + \psi_{y} \psi)$.  The former combination 
can be further reduced using the identity
\be
\bar \psi_{y} \bar \psi - \psi_{y} \psi = \partial_{y}(\bar \psi \psi)
- 2 {\partial \ln f \over \partial \phi}\phi_{y} \bar \psi \psi \,.
\ee
However, the latter combination cannot be so reduced, and thus, its 
coefficient must vanish. After some computation, we find that the 
total contribution of such terms in (\ref{integconstraint}) is
\be
(\bar \psi_{y} \bar \psi + \psi_{y} \psi) \phi_{y} \left\{
-{3\over 8} \left[\alpha \sin{1\over 2}(\phi - \phi_{0}) 
- 4 \sin {\phi\over 2}\right] 
+ 12 f^{2} {\partial \ln f \over \partial \phi}
\right\} \,.
\label{mustgo}
\ee
Recalling the result (\ref{fresult}) for $f(\phi)$, we find that 
the expression in (\ref{mustgo}) within braces does vanish, 
provided that the integration constant $C$ is given by \footnote{There
is a second (negative) root, which is ruled out by the requirement
that $f$ be real.}
\be
C = \sqrt{\alpha^{2} - 8 \alpha \cos {\phi_{0}\over 2} + 16} \,.
\label{Cresult}
\ee
The expression for $f(\phi)$ then takes the simplified form
\be
f(\phi) = {\sqrt{C}\over 2}\sin{1\over 4}(\phi - D) \,, \qquad 
\mbox{where} \quad 
\tan {D\over 2} ={\alpha \sin{\phi_{0}\over 2}\over
\alpha \cos{\phi_{0}\over 2} -4} \,.
\label{simpler}
\ee 

After further computation,  we find that (\ref{integconstraint}) 
is given by
\be
\lefteqn{i\bar \psi \psi \Bigg\{ \phi_{y} \Bigg[ 
{3\over 2} \sin \phi - {3\over 4} \alpha \sin {\phi_{0}\over 2} - 
{3\over 32}\alpha^{2} \sin (\phi - \phi_{0}) 
- 6 f^{2} \sin {\phi\over 2}  } \non \\
& &+ \left({3\over 8} \alpha^{2} \sin^{2} {1\over 2}(\phi - \phi_{0}) 
+ 6 \cos \phi - 6 \cos^{2}{\phi\over 2} - 24 f^{2} \right)
{\partial \ln f \over \partial \phi} 
\Bigg] \non  \\
& &+ 2\phi_{y}^{3} \left[ {\partial^{3} \ln f \over \partial \phi^{3}}
+ 3 {\partial \ln f \over \partial \phi} 
{\partial^{2} \ln f \over \partial \phi^{2}} 
+ {1\over 16} {\partial \ln f \over \partial \phi} 
+ \left({\partial \ln f \over \partial \phi}\right)^{3} \right]  \non \\
& &+ 2 \phi_{y}\phi_{yy}\left[ {3\over 16} 
+ 3 {\partial^{2} \ln f \over \partial \phi^{2}}
+ 3 \left({\partial \ln f \over \partial \phi}\right)^{2} \right]
\Bigg\} \,,
\ee
up to a total $y$-derivative.  Remarkably, upon substituting the
result (\ref{simpler}) for $f(\phi)$, we find that the quantities in
each of the square brackets vanish.  We conclude that $P_{3}$ is
indeed an integral of motion.

Finally, we remark that the boundary model also has the integral of 
motion
\be
P'_{s}= \int_{-\infty}^{0}dx\ (T_{s+1} - \overline{T}_{s+1} + 
\Theta_{s-1} - \overline{\Theta}_{s-1}) - i\Sigma'_{s}(y) \,,
\label{iomprime}
\ee
provided that the constraint
\be
T_{s+1} + \overline{T}_{s+1} - \Theta_{s-1} - \overline{\Theta}_{s-1}
= \partial_{y} \Sigma'_{s}(y) 
\label{constraintprime}
\ee 
is satisfied at $x=0$.  We expect that the boundary SSG model has (for 
$s={1\over 2}$, at least) such integrals of motion if a different set 
of boundary conditions is imposed, corresponding to a boundary 
Lagrangian of the form
\be
{\cal L}'_{b} = -\bar \psi \psi + i a \partial_{y} a 
- 2 i f(\phi) a (\psi + \bar \psi) + {\cal B}(\phi) \,.
\label{boundLprime}
\ee
However, we shall not pursue those calculations here.

\section{Discussion}\label{sec:discuss}

We have shown that the boundary SSG model with boundary Lagrangian 
(\ref{boundL}), with ${\cal B}(\phi)$ given by (\ref{Bpotential}) and 
$f(\phi)$ given by (\ref{simpler}), has the integrals 
of motion $P_{s}$ (\ref{iom}) with $s={1\over 2}$ and $s=3$.  The fact 
that $P_{1\over 2}$ is an integral of motion means that the model has 
on-shell supersymmetry.  The fact that $P_{3}$ is an integral of 
motion is strong evidence that the model is integrable.  We emphasize 
that this boundary interaction, just like the one for the (Bosonic) 
boundary SSG model, has two continuous parameters, $\alpha$ and 
$\phi_{0}$.  As mentioned in the Introduction, we were 
motivated to look for such boundary interactions by our earlier 
results \cite{AN2} on the boundary SYL model.

We conjecture that the boundary SSG model is integrable, and that the 
corresponding boundary $S$ matrix $\SSS(\theta)$ for the $n=1$ 
breather supermultiplet is given by \cite{AN1}
\be
\SSS(\theta)= \SSS_{SG}(\theta \,; \eta \,, \vartheta )\ 
\SSS_{SUSY}^{(\varepsilon=+1)}(\theta \,; \varphi ) \,,
\label{Smatrix}
\ee
where $ \SSS_{SG}(\theta \,; \eta \,, \vartheta )$ is the $n=1$ 
breather boundary SG reflection factor \cite{Gh}, and 
$\SSS_{SUSY}^{(\varepsilon)}(\theta \,; \varphi )$ is a $2 \times 2$ 
matrix given in \cite{AK, AN1}. Indeed, $\SSS(\theta)$ satisfies the 
boundary Yang-Baxter equation, as well as unitarity and boundary 
cross-unitarity \cite{GZ}. Moreover, $\SSS(\theta)$ commutes with a 
supersymmetry charge corresponding to the integral of motion 
$P_{1\over 2}$.\footnote{The matrix 
$\SSS_{SUSY}^{(\varepsilon)}(\theta \,; \varphi )$ with 
$\varepsilon=-1$ commutes with a supersymmetry charge corresponding to 
$P'_{s}$ (\ref{iomprime}) with $s={1\over 2}$.} The proof is 
essentially the same as the one given in \cite{AN2} for the boundary 
SYL model. Note that the boundary term $\Sigma_{1\over 2}$ 
(\ref{boundterm}) corresponds (up to a proportionality factor) to 
$(-1)^{F}$, where $F$ denotes Fermion number.

Although the $S$ matrix (\ref{Smatrix}) involves three boundary 
parameters $\eta$, $\vartheta$, $\varphi$, we expect that there is 
one relation among these parameters, so that only two parameters are 
independent, in concordance with the boundary action. Indeed, a 
parallel situation occurs for the boundary SYL model \cite{AN2}, whose 
boundary $S$ matrix nominally involves two boundary parameters, but 
only one of them is independent. For the boundary SSG model, the 
relation among the parameters can presumably be found in the same 
manner as for the boundary SYL model \cite{AN2}: namely, by imposing 
the constraint \cite{GZ} that near a pole $i v^{\alpha}_{0 a}$ of the 
boundary $S$ matrix associated with the excited boundary state 
$|\alpha \rangle_{B}$ (which can be interpreted as a boundary bound 
state of particle $A_{a}$ with the boundary ground state 
$|0\rangle_{B}$), the boundary $S$ matrix must have the form
\be
\SSS_{a}^{b}(\theta) \simeq {i\over 2}
{g^{\alpha}_{a 0} g^{b 0}_{\alpha}\over \theta - i v^{\alpha}_{0 a}} \,,
\ee 
where $g^{\alpha}_{a 0}$ are boundary-particle couplings.  It would be 
interesting to explicitly work out this relation for the boundary SSG 
model, and to find the precise relation of the two independent 
parameters of the boundary $S$ matrix to the parameters $\alpha$, 
$\phi_{0}$ of the boundary action.  The latter problem has already 
been addressed for the case of the (Bosonic) boundary SG model 
\cite{AlZ, Co}, and we expect that similar techniques can be applied 
to the supersymmetric case. Moreover, analogous results should also 
hold for boundary $S$ matrices for the SSG higher breathers and 
solitons.

\section*{Acknowledgments}

I thank C. Ahn for his collaboration at an early stage of this work,
and O. Alvarez and J. S\'anchez-Guill\'en for a valuable comment. 
This work was supported in part by the National Science Foundation
under Grant PHY-9870101.

\bigskip

\noindent
{\bf Note added:}

We have verified that the boundary SSG model with the boundary
Lagrangian ${\cal L}'_{b}$ (\ref{boundLprime}) indeed has the
integrals of motion $P'_{1\over 2}$ (\ref{iomprime}) and $P_{3}$
(\ref{iom}), with the same ${\cal B}(\phi)$ (\ref{Bpotential}), 
and $f(\phi) = {\sqrt{C}\over 2}\cos{1\over 4}(\phi - D)$ where
$C$ and $D$ are given by (\ref{Cresult}) and (\ref{simpler}) with 
$\alpha \rightarrow -\alpha$. In the conformal limit, the 
boundary Lagrangians (\ref{boundL}) and (\ref{boundLprime}) give 
Neveu-Schwarz and Ramond boundary conditions, respectively.

\end{document}